\documentclass[conference]{IEEEtran}
\IEEEoverridecommandlockouts

\usepackage{cite}
\usepackage{amsmath,amssymb,amsfonts}
\usepackage{graphicx}
\usepackage{adjustbox}
\usepackage{url}
\usepackage{booktabs}
\usepackage{xcolor}
\usepackage{tikz}
\usetikzlibrary{arrows.meta,positioning,fit,backgrounds,calc,shapes.geometric}
\usepackage{pgfplots}
\pgfplotsset{compat=1.17}
\usepackage{etoolbox}    
\usepackage[hidelinks]{hyperref}
\AtBeginEnvironment{figure}{\vspace*{8pt}}    

\tikzset{
  edgebox/.style   ={rectangle, rounded corners=2pt, draw=black!70,
                     fill=blue!8, minimum height=7mm, minimum width=15mm,
                     align=center, font=\scriptsize},
  brokerbox/.style ={rectangle, rounded corners=2pt, draw=black!70,
                     fill=orange!15, minimum height=7mm, minimum width=15mm,
                     align=center, font=\scriptsize},
  cloudbox/.style  ={rectangle, rounded corners=2pt, draw=black!70,
                     fill=green!10, minimum height=7mm, minimum width=15mm,
                     align=center, font=\scriptsize},
  plantbox/.style  ={rectangle, rounded corners=2pt, draw=black!70,
                     fill=gray!12, minimum height=7mm, minimum width=15mm,
                     align=center, font=\scriptsize},
  dataflow/.style  ={-{Stealth[length=2mm]}, thick, draw=black!75},
  ctrlflow/.style  ={-{Stealth[length=2mm]}, thick, draw=red!70, dashed},
  grouplbl/.style  ={font=\scriptsize\bfseries, text=black!60},
}
\begin{document}

\title{A Kafka-Centric Communication Fabric for Near-Real-Time,\\
Cloud-Replicated Closed-Loop Manufacturing Process Control}

\author{
\IEEEauthorblockN{Zhengyang ``Cissy'' Gu}
\IEEEauthorblockA{\textit{Data Science}\\
\textit{Liveline Technologies}\\
Livonia, MI, USA\\
cissy.gu@liveline.tech}
\and
\IEEEauthorblockN{John Burtenshaw}
\IEEEauthorblockA{\textit{Engineering}\\
\textit{Liveline Technologies}\\
Livonia, MI, USA\\
john.burtenshaw@liveline.tech}
\and
\IEEEauthorblockN{Joseph E. Hernandez}
\IEEEauthorblockA{\textit{Engineering}\\
\textit{Liveline Technologies}\\
Livonia, MI, USA\\
joseph.hernandez@liveline.tech}
\and
\IEEEauthorblockN{Chris Couch}
\IEEEauthorblockA{\textit{Chief Executive Officer}\\
\textit{Liveline Technologies}\\
Livonia, MI, USA\\
chris.couch@liveline.tech}
\thanks{\normalfont{This project was funded by Liveline Technologies. Liveline Technologies is dedicated to improving manufacturing performance by harnessing the power of Artificial Intelligence to automate complex processes and predict future problems.}}
\thanks{\normalfont{This is the author's version of this work. It has been accepted for publication in the 2026 IEEE Real Time Communications Conference and Expo. The final published version will be made available on IEEE Xplore.}}
\thanks{\normalfont{\copyright~2026 IEEE. Personal use of this material is permitted. Permission from IEEE must be obtained for all other uses, in any current or future media, including reprinting/republishing this material for advertising or promotional purposes, creating new collective works, for resale or redistribution to servers or lists, or reuse of any copyrighted component of this work in other works.}}

}

\maketitle

\begin{abstract}
Smart manufacturing needs to move sensor data off the plant floor, react to it,
and feed decisions back to actuators within a bounded time. Programmable logic
controllers (PLCs) handle fast, deterministic, safety-critical actuation well,
but they are not designed for the higher-level functions that Industry~4.0
demands, such as predictive maintenance, machine-learning inference, and
cross-facility analytics. Those functions need a communication substrate that is
scalable, durable, and observable. We present the communication architecture of
a production system that provides such a substrate and closes the loop back to
the plant in near real time. The design is built entirely from industry-standard
components: Apache Kafka as the streaming backbone, OPC-UA for PLC connectivity,
a relational time-series database and a metrics store for persistence, and JSON
for serialization. The novelty is architectural. We show how these standard
pieces are stitched together for closed-loop industrial control through four
design decisions: a single per-line event stream that simultaneously drives
control, monitoring, machine learning, and durable recording, so that one data
producer serves many independent consumers; a protocol bridge that turns polled
OPC-UA traffic into publish/subscribe streaming, aligns per-signal timestamps
onto a common time base to remove cross-signal jitter, and provides a symmetric
actuation path; a transport technique that carries sub-second process dynamics
over a coarser publication cadence by packing timestamped samples into
fixed-order arrays; and an edge-to-cloud replication scheme in which the edge
stays authoritative, so local control keeps running during wide-area-network
(WAN) outages while cloud analytics operate on replicated data. We give the
latency budget of the loop, report a measured broker-transport latency, and
describe our operational experience. We are explicit that the system provides
soft, near-real-time behavior rather than hard-real-time guarantees.
\end{abstract}

\begin{IEEEkeywords}
real-time communication, publish/subscribe, Apache Kafka, OPC-UA, industrial
IoT, edge computing, closed-loop control, time-series data, stream processing,
smart manufacturing.
\end{IEEEkeywords}

\section{Introduction}

Manufacturing process control has long lived inside the deterministic world of
the programmable logic controller (PLC), the ruggedized industrial computer that
drives machinery. A PLC excels at a specific class of functions: reading sensors
and driving actuators on a fixed millisecond cycle, enforcing safety interlocks,
and sequencing equipment through deterministic, hard-real-time logic. A
manufacturing \emph{line}, the sequence of machines that together produce a
product, may contain several PLCs coordinating in this way.

What a PLC does not do is the higher-level work that Industry~4.0 and smart
manufacturing increasingly require~\cite{lasi2014,lee2015,zhong2017}. It does not
aggregate data across an entire line, retain history, run statistical or
machine-learning (ML) models, or compare the current production against past
production. These are exactly the functions that drive modern quality and yield:
predictive maintenance, which forecasts equipment degradation before it causes
scrap or downtime~\cite{icad2026}; anomaly and defect detection; deep-learning
prediction used to augment or stand in for physical measurements, so that a
model acts as a low-cost ``virtual sensor'' alongside real instrumentation; and
closed-loop optimization that adjusts machine settings toward a target. The same
family of AI functions appears widely across industry, including
computer-vision surface inspection, soft sensing of quantities that are hard to
measure directly, remaining-useful-life estimation, and data-driven process
optimization, and it is a step toward the closed-loop digital twin~\cite{tao2019}.

These higher-level functions want a communication substrate that the PLC network
was never meant to provide: one that is horizontally scalable, durable,
observable, and friendly to general-purpose compute. Surveys of industrial
communication identify this substrate, rather than the analytics on top of it,
as the recurring bottleneck~\cite{wollschlaeger2017,vitturi2019}, and edge and
cloud computing research argues for placing latency-critical computation close to
the source~\cite{shi2016,satya2017}.

This paper describes the communication fabric of a system deployed in production
across multiple manufacturing facilities that provides that substrate and, in
addition, closes the loop back to the plant. Sensor and equipment data are
lifted off PLCs by way of OPC-UA, published onto an Apache Kafka cluster, and
consumed at the same time by services that can compute new machine settings and
write them back to the PLCs, by a process monitor and ML inference services, and
by a service that durably records every measurement. The same data is replicated
to the cloud, where a second recorder and further analytics run.

The protocols and brokers we use are industry standards; we introduce no new
wire format. Our contribution is the architecture, specifically the way these
standard pieces are stitched together to satisfy near-real-time closed-loop
control, together with the design decisions and trade-offs that make the
composition work in production. Our contributions are:

\begin{itemize}
  \item A \textbf{single-stream, many-consumer} fabric in which one published
  data stream per line serves control, monitoring, ML, and recording at once
  (Section~\ref{sec:arch}).
  \item A \textbf{protocol bridge} that converts polled OPC-UA traffic into
  publish/subscribe streaming, \textbf{aligns per-signal timestamps} onto a
  common time base so a multi-signal snapshot is jitter-free, and provides a
  symmetric actuation path (Section~\ref{sec:acq}).
  \item A \textbf{high-rate transport technique} that carries sub-second (for
  example 250\,ms) process dynamics over a coarser one-second publication cadence
  by packing timestamped samples into fixed-order arrays, so each consumer may
  take either an aggregate or the full-resolution series
  (Section~\ref{sec:highrate}).
  \item An \textbf{edge-to-cloud replication} design that keeps the edge
  authoritative for job identity and line state, so local control keeps running
  during WAN outages while cloud analytics work on replicated data
  (Section~\ref{sec:replication}).
\end{itemize}

The system targets \emph{near-real-time}, soft real time: an end-to-end
control-loop budget of one to two seconds (Section~\ref{sec:latency}), not the
microsecond determinism of the fieldbus. The governing rule is that the system
must never act on data more than about two seconds old. The hard-real-time
interlocks stay in the PLC. Section~\ref{sec:discussion} discusses this boundary.

\section{Background and Related Work}

We frame the problem with three questions: how do machines speak today, why add a
streaming layer, and how do the pieces fit into a plant.

\subsection{How machines speak: OPC-UA and the fieldbus}
OPC-UA (IEC~62541)~\cite{opcua} is the common, vendor-neutral language for
industrial data. It is capable and secure, with browsable data models,
subscriptions, and encrypted sessions. Its limitation, for our purpose, is its
client-server shape: it is built for one client to ask a server for data, not
for one data source to feed many independent, changing analytics consumers at
once. Studies of industrial protocols report that a single OPC-UA server slows
down as the number of concurrent clients grows~\cite{profanter2019}. Below
OPC-UA, at the wire level, Time-Sensitive Networking (TSN)~\cite{tsn} and OPC-UA
PubSub over TSN~\cite{bruckner2019} deliver hard-real-time determinism. They are
complementary to the plant-to-cloud fabric we describe, not competitors: they
make the fast reflexes deterministic, while our fabric carries the slower,
higher-level loop.

\subsection{Why a streaming layer: the durable log}
Apache Kafka~\cite{kafka} popularized a simple idea that fits our problem well:
treat the data stream as a durable, replayable log that many consumers read
independently at their own pace. This log-centric style is now a common backbone
for data-intensive systems~\cite{kleppmann2017} and stream
processing~\cite{akidau2015,carbone2015}. It contrasts with lighter message
buses. RabbitMQ and AMQP~\cite{amqp} and the lightweight IoT protocol
MQTT~\cite{mqtt} push messages to subscribers and typically keep no replayable
history; empirical comparisons quantify their latency, throughput, and fan-out
behavior against Kafka~\cite{dobbelaere2017,naik2017,luzuriaga2015}. We return to
why durability and replay matter for us in Section~\ref{sec:discussion}.

\subsection{How the pieces fit a plant: edge, cloud, and consistency}
Reference models for factory data, such as the ISA-95 / IEC~62264
hierarchy~\cite{isa95} and the ``unified namespace'' idea, advocate decoupling
shop-floor data through a central broker. Our system is a concrete instance of
that idea that also closes the loop to actuators and keeps running when the link
to the cloud fails. Choosing local availability over global consistency during
such a partition is the classic trade stated by the CAP theorem~\cite{cap}: when
the network splits, a system can stay available or stay globally consistent, but
not both, and for a control loop the right choice is to stay available at the
edge.

\section{System Architecture}
\label{sec:arch}

The architecture is organized around the manufacturing line as the unit of
deployment, and around one shared event stream per line as the unit of
communication. Each line runs an isolated set of cooperating microservices on an
edge compute node, all communicating with a Kafka cluster. The shared software
base is a microservice framework we refer to as the digital control platform
(DCP), which provides a common service skeleton, Kafka integration, signal
management, an embedded HTTP interface, and metrics.

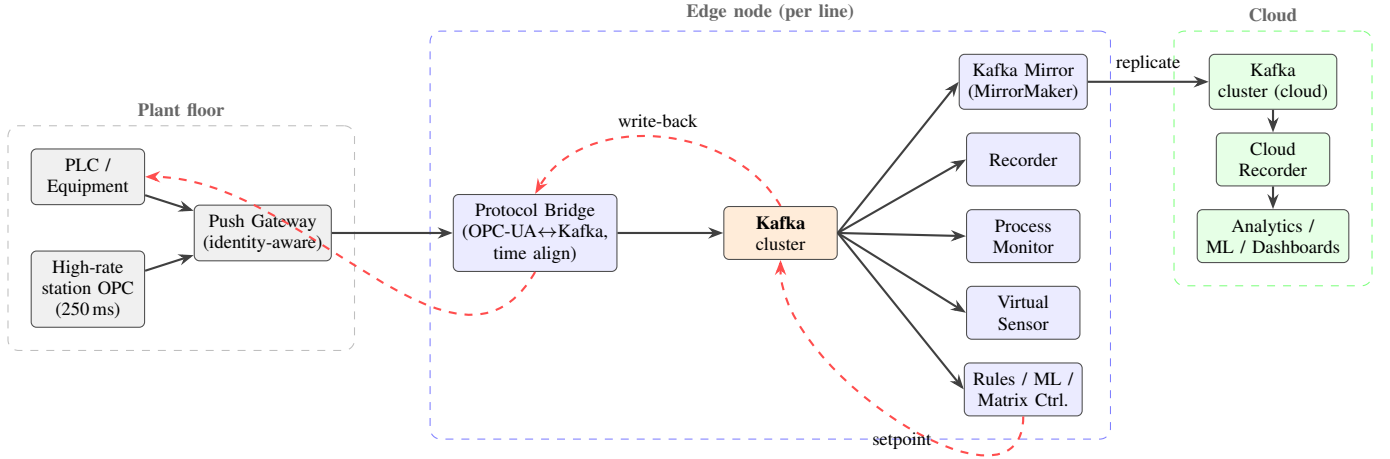
\begin{figure*}[t]
\centering
\adjustbox{max width=\textwidth}{%
\begin{tikzpicture}[node distance=6mm and 10mm]
\node[plantbox] (plc) {PLC /\\ Equipment};
\node[plantbox, below=of plc] (afl) {High-rate\\station OPC\\(250\,ms)};
\node[plantbox, right=14mm of $(plc)!0.5!(afl)$] (gw)
     {Push Gateway\\(identity-aware)};
\node[edgebox, right=16mm of gw] (bridge) {Protocol Bridge\\(OPC-UA$\leftrightarrow$Kafka,\\time align)};
\node[brokerbox, right=14mm of bridge] (kafka)
     {\textbf{Kafka}\\cluster};
\node[edgebox, right=16mm of kafka, yshift=20mm] (mirror) {Kafka Mirror\\(MirrorMaker)};
\node[edgebox, below=3mm of mirror] (hist) {Recorder};
\node[edgebox, below=3mm of hist] (pmon) {Process\\Monitor};
\node[edgebox, below=3mm of pmon] (vsen) {Virtual\\Sensor};
\node[edgebox, below=3mm of vsen] (ctrl) {Rules / ML /\\Matrix Ctrl.};
\node[cloudbox, right=16mm of mirror] (ckafka) {Kafka\\cluster (cloud)};
\node[cloudbox, below=3mm of ckafka] (chist) {Cloud\\Recorder};
\node[cloudbox, below=3mm of chist] (canalytics) {Analytics /\\ML / Dashboards};

\draw[dataflow] (plc) -- (gw);
\draw[dataflow] (afl) -- (gw);
\draw[dataflow] (gw) -- (bridge);
\draw[dataflow] (bridge) -- (kafka);
\draw[dataflow] (kafka.east) -- (mirror.west);
\draw[dataflow] (kafka.east) -- (hist.west);
\draw[dataflow] (kafka.east) -- (pmon.west);
\draw[dataflow] (kafka.east) -- (vsen.west);
\draw[dataflow] (kafka.east) -- (ctrl.west);
\draw[dataflow] (mirror) -- node[above,font=\scriptsize]{replicate} (ckafka);
\draw[dataflow] (ckafka.south) -- (chist.north);
\draw[dataflow] (chist.south) -- (canalytics.north);
\draw[ctrlflow] (ctrl.south) to[out=-90,in=-90] node[below,font=\scriptsize,pos=0.5]{setpoint} (kafka.south);
\draw[ctrlflow] (kafka.north) to[out=120,in=60] node[above,font=\scriptsize,pos=0.5]{write-back} (bridge.north);
\draw[ctrlflow] (bridge.south) to[out=-120,in=0] (plc.east);

\begin{scope}[on background layer]
  \node[draw=gray!50, dashed, rounded corners, fit=(plc)(afl)(gw),
        inner sep=3mm, label={[grouplbl]above:Plant floor}] {};
  \node[draw=blue!50, dashed, rounded corners,
        fit=(bridge)(kafka)(mirror)(hist)(pmon)(vsen)(ctrl),
        inner sep=3mm, label={[grouplbl]above:Edge node (per line)}] {};
  \node[draw=green!50, dashed, rounded corners, fit=(ckafka)(chist)(canalytics),
        inner sep=3mm, label={[grouplbl]above:Cloud}] {};
\end{scope}
\end{tikzpicture}%
}
\caption{System architecture. Solid arrows are the data path (plant to bridge to
Kafka to many independent consumers). A dedicated mirror service (MirrorMaker)
replicates the edge cluster to an independent cloud cluster; dashed arrows are
the closed-loop control write-back (controller to Kafka to bridge to PLC). The
bridge is the only component that speaks OPC-UA; every other service speaks only
Kafka. The cloud is fed by the mirror and is never in the control path.}
\label{fig:arch}
\end{figure*}

Figure~\ref{fig:arch} shows the components and data flows. The services on a
line are: the \textbf{Protocol Bridge}, which connects to PLCs by way of OPC-UA
and is the only component that speaks the industrial protocol directly; the
\textbf{Recorder}, which durably stores the equipment data stream
(Section~\ref{sec:recorder}); the \textbf{Process Monitor}, which computes
on-line statistics and process health; the \textbf{Virtual Sensor}, which
publishes ML predictions as if they were physical signals; and the
\textbf{Rules, ML, and Matrix Controllers}, which compute new machine settings by
rule-based, ML-based, and multivariable strategies and write them back toward the
PLC.

\subsection{One stream, many consumers}
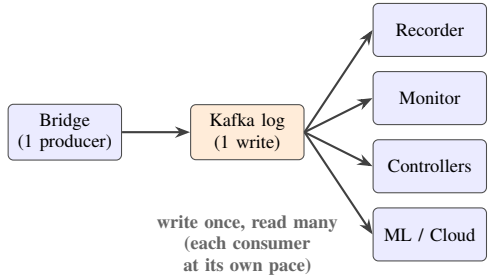
\begin{figure}[t]
\centering
\adjustbox{max width=\columnwidth}{%
\begin{tikzpicture}[node distance=4mm and 9mm, font=\scriptsize]
\node[edgebox] (b) {Bridge\\(1 producer)};
\node[brokerbox, right=of b] (k) {Kafka log\\(1 write)};
\node[edgebox, right=of k, yshift=13.5mm] (c1) {Recorder};
\node[edgebox, right=of k, yshift=4.5mm]  (c2) {Monitor};
\node[edgebox, right=of k, yshift=-4.5mm] (c3) {Controllers};
\node[edgebox, right=of k, yshift=-13.5mm](c4) {ML / Cloud};
\draw[dataflow] (b) -- (k);
\foreach \c in {c1,c2,c3,c4}{ \draw[dataflow] (k.east) -- (\c.west); }
\node[grouplbl, below=6mm of k, text width=3.2cm, align=center]
     {write once, read many\\(each consumer at its own pace)};
\end{tikzpicture}%
}
\caption{The single-stream, many-consumer principle. One producer writes to the
durable log; independent consumers each read the whole stream at their own rate,
so adding or restarting a consumer never disturbs the others.}
\label{fig:fanout}
\end{figure}

The keystone of the design is that the bridge publishes equipment data
\emph{once} to Kafka, and every other service reads it independently
(Fig.~\ref{fig:fanout}). There is no direct coupling between the data source and
its consumers. A new analytics service is added simply by subscribing to Kafka,
with no change to the producer or to existing consumers. Because Kafka keeps the
log, a starting or recovering consumer can rewind and replay history; the digital control platform (DCP) uses this to warm up a service's working buffers from past data before it goes live. This one decision lets the same bytes serve both a latency-sensitive controller and a throughput-oriented recorder, each reading at its own pace.

\subsection{Data model and serialization}
Messages are serialized as JSON. A reading carries a millisecond timestamp, a
numeric or structured value, and an explicit data-quality flag, and a shared
data-model library keeps producers and consumers in agreement on the format. We
use JSON rather than a binary schema format because its readability and tooling
have outweighed the size cost at our message rates; Section~\ref{sec:discussion}
revisits this.

\section{Data Acquisition and Actuation}
\label{sec:acq}

This section describes how data enters the stream from the plant and how control
decisions leave the stream to reach the plant. Both directions pass through one
component, the Protocol Bridge, which is the single place that speaks OPC-UA.

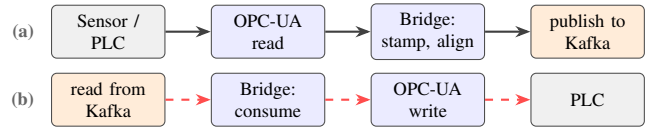
\begin{figure}[t]
\centering
\adjustbox{max width=\columnwidth}{%
\begin{tikzpicture}[node distance=3.5mm and 6mm, font=\scriptsize]
\node[plantbox] (r1) {Sensor /\\PLC};
\node[edgebox, right=of r1] (r2) {OPC-UA\\read};
\node[edgebox, right=of r2] (r3) {Bridge:\\stamp, align};
\node[brokerbox, right=of r3] (r4) {publish to\\Kafka};
\draw[dataflow] (r1) -- (r2); \draw[dataflow] (r2) -- (r3); \draw[dataflow] (r3) -- (r4);
\node[grouplbl, left=1mm of r1] {(a)};
\node[brokerbox, below=9mm of r1.west, anchor=west] (w1) {read from\\Kafka};
\node[edgebox, right=of w1] (w2) {Bridge:\\consume};
\node[edgebox, right=of w2] (w3) {OPC-UA\\write};
\node[plantbox, right=of w3] (w4) {PLC};
\draw[ctrlflow] (w1) -- (w2); \draw[ctrlflow] (w2) -- (w3); \draw[ctrlflow] (w3) -- (w4);
\node[grouplbl, left=1mm of w1] {(b)};
\end{tikzpicture}%
}
\caption{The two directions through the bridge. (a) Read path, plant to stream:
the bridge reads OPC-UA nodes, stamps and time-aligns each reading, and publishes
to Kafka. (b) Write path, stream to plant: the bridge consumes a controller's
setpoint from Kafka and issues the OPC-UA write to the PLC.}
\label{fig:paths}
\end{figure}

The bridge maintains an OPC-UA session to the PLC, with support for encrypted
sessions and certificate- or password-based authentication, and on the other
side it is an ordinary Kafka client. Figure~\ref{fig:paths} shows the two
directions.

\subsection{Read path: plant to stream}
The bridge samples or subscribes to OPC-UA nodes and republishes each reading to
Kafka, with its source timestamp and quality. This turns the polled,
session-oriented OPC-UA model into an append-only stream that any number of
consumers can read without adding load to the PLC. When data originates in a
separate OPC server, for example a high-rate inline station, it is forwarded to
our OPC layer over a push-based, identity-aware gateway, which we prefer to
polling for freshness; this leg can add up to about one second.

\subsection{Write path: stream to plant}
Control is symmetric. A controller that has computed a new setpoint, the target
value it wants a process variable to reach, publishes it to Kafka. The bridge
consumes it from Kafka and issues the OPC-UA write to the PLC. The controllers
therefore never speak OPC-UA; they speak Kafka, and the bridge is the single,
auditable point through which all actuation flows. This buys three things: every
commanded setpoint is durably logged for later replay, controllers can be built
and tested entirely against the stream, and the safety-critical OPC interface has
exactly one implementation to harden. The hard-real-time interlocks and limits
stay inside the PLC; the loop only adjusts setpoints within them.

\subsection{Timestamp alignment and jitter management}
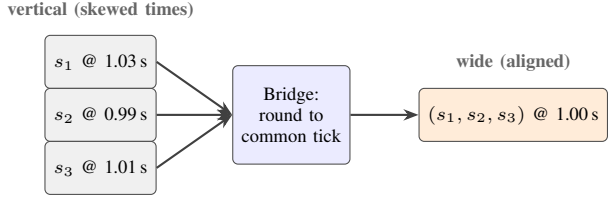
\begin{figure}[t]
\centering
\adjustbox{max width=\columnwidth}{%
\begin{tikzpicture}[font=\scriptsize, node distance=2mm]
\node[plantbox, minimum width=13mm] (s1) at (0,1.4) {$s_1$ @ 1.03\,s};
\node[plantbox, minimum width=13mm] (s2) at (0,0.7) {$s_2$ @ 0.99\,s};
\node[plantbox, minimum width=13mm] (s3) at (0,0.0) {$s_3$ @ 1.01\,s};
\node[grouplbl, above=1mm of s1] {vertical (skewed times)};
\node[edgebox, right=10mm of s2, minimum height=13mm] (br) {Bridge:\\round to\\common tick};
\draw[dataflow] (s1.east) -- (br.west);
\draw[dataflow] (s2.east) -- (br.west);
\draw[dataflow] (s3.east) -- (br.west);
\node[brokerbox, right=9mm of br, minimum width=20mm] (out)
     {$(s_1,s_2,s_3)$ @ 1.00\,s};
\node[grouplbl, above=1mm of out] {wide (aligned)};
\draw[dataflow] (br.east) -- (out.west);
\end{tikzpicture}%
}
\caption{Timestamp alignment. Signals arriving with slightly different source
times are rounded by the bridge onto a common tick, producing a jitter-free,
co-timed snapshot for consumers.}
\label{fig:align}
\end{figure}

Signals arrive one value at a time, each with its own source time, but consumers
want a coherent snapshot in which co-occurring signals share one timestamp
(Fig.~\ref{fig:align}). The bridge performs this reconciliation. It aligns
readings onto a common, rounded time base, a line clock synchronized by NTP or
PTP~\cite{ntp,ptp}, so that a snapshot for a given instant is exactly co-timed
and free of cross-signal skew. Correctness rests on one principle: time is
stamped in the OPC layer, at or near the source, while the data is still live. As
long as the OPC layer stamps correctly, every consumer downstream can trust the
time. For values that pass through a foreign OPC server, true fidelity would
require that server to stamp at its own OPC layer; absent that, the bridge's
rounding bounds the residual skew. This alignment is what lets a controller act
on a coherent multi-signal snapshot rather than on values that are each fresh but
mutually misaligned in time.

\section{Transporting High-Rate Signals}
\label{sec:highrate}

This section addresses a recurring tension in plant-to-cloud communication: the
transport cadence is often coarser than the process dynamics. In our deployments
the practical sampling and publication cadence is one second, yet some processes
evolve faster; a high-rate inline station is meaningful at 250\,ms resolution.
Simply down-sampling to the transport rate would discard most of the signal and
blur the timing of transient events.

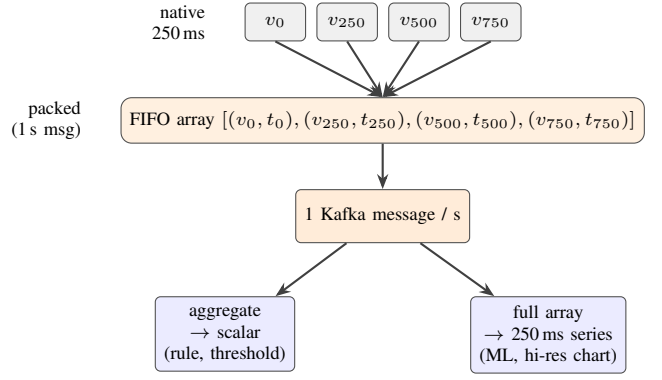
\begin{figure}[t]
\centering
\vspace*{9pt}
\adjustbox{max width=\columnwidth}{%
\begin{tikzpicture}[font=\scriptsize, node distance=2mm]
\foreach \i/\t in {0/0,1/250,2/500,3/750} {
  \node[plantbox, minimum width=8mm, minimum height=5mm]
        (n\i) at (\i*0.95,2) {$v_{\t}$};
}
\node[left=4mm of n0, align=right] {native\\250\,ms};
\node[draw=black!70, fill=orange!12, rounded corners, minimum width=3.9cm,
      minimum height=6mm, align=center] (arr) at (1.42,0.7)
      {FIFO array $[(v_0,t_0),(v_{250},t_{250}),(v_{500},t_{500}),(v_{750},t_{750})]$};
\node[left=4mm of arr, align=right] {packed\\(1\,s msg)};
\foreach \i in {0,1,2,3} { \draw[dataflow] (n\i.south) -- (arr.north); }
\node[brokerbox, below=6mm of arr, minimum width=2.2cm] (msg) {1 Kafka message / s};
\draw[dataflow] (arr) -- (msg);
\node[edgebox, below left=7mm and 0mm of msg, align=center] (agg)
     {aggregate\\$\rightarrow$ scalar\\(rule, threshold)};
\node[edgebox, below right=7mm and 0mm of msg, align=center] (full)
     {full array\\$\rightarrow$ 250\,ms series\\(ML, hi-res chart)};
\draw[dataflow] (msg) -- (agg);
\draw[dataflow] (msg) -- (full);
\end{tikzpicture}%
}
\caption{High-rate transport. Four 250\,ms samples, each with its own source
timestamp, are packed into a FIFO array on the PLC and shipped as one message per
second. One message serves both a low-resolution consumer (which averages to a
scalar) and a high-resolution consumer (which reads the full series).}
\label{fig:highrate}
\end{figure}

Our solution decouples \emph{sampling resolution} from \emph{transport cadence}
(Fig.~\ref{fig:highrate}). At the PLC, high-rate samples are captured at their
native resolution and written, each with its own timestamp, into a
fixed-capacity, first-in-first-out (FIFO) array. Once per transport interval the
whole array is shipped as a single message. Because every element is stamped in
the OPC layer, no timing is lost even though the array is delivered on the slower
cadence, and the FIFO discipline guarantees samples are consumed in the exact
order and with the exact times they were produced. On the consumer side, the digital control platform wraps the array so a service chooses its own view: a consumer that needs one scalar, for example a rule checking a threshold, receives the array aggregated to a single value, while a consumer that needs fidelity, for example an ML model or a high-resolution chart, receives the full array. One published message thus serves both a low-resolution and a high-resolution consumer without duplicating traffic, which is the single-stream, many-consumer principle applied to the time dimension.

This technique is the clearest example in the system of a media-transport
optimization: it preserves sub-interval temporal fidelity and per-sample timing
across a constrained link, and it manages jitter by moving time-stamping to the
source. It generalizes to any setting where a fixed publication cadence must
carry a faster-evolving signal.

\section{Edge-to-Cloud Replication}
\label{sec:replication}

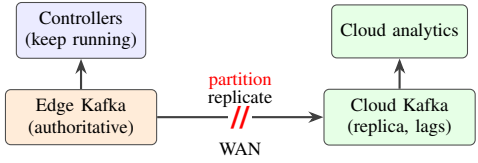
\begin{figure}[t]
\centering
\vspace*{9pt}
\adjustbox{max width=\columnwidth}{%
\begin{tikzpicture}[font=\scriptsize, node distance=5mm and 12mm]
\node[brokerbox, minimum width=20mm] (ek) {Edge Kafka\\(authoritative)};
\node[edgebox, above=4mm of ek] (ec) {Controllers\\(keep running)};
\node[cloudbox, right=22mm of ek, minimum width=20mm] (ck) {Cloud Kafka\\(replica, lags)};
\node[cloudbox, above=4mm of ck] (ca) {Cloud analytics};
\draw[dataflow] (ek) -- (ec);
\draw[dataflow] (ck) -- (ca);
\draw[dataflow] (ek) -- node[above,align=center]{replicate} node[below,yshift=-2mm]{WAN} (ck);
\node[red, font=\bfseries\large, fill=white, inner sep=1pt] (x) at ($(ek)!0.5!(ck)$) {//};
\node[red, font=\scriptsize, above=0.5mm of x] {partition};
\end{tikzpicture}%
}
\caption{Edge-to-cloud replication. The edge is authoritative and keeps the
control loop running on local Kafka. If the WAN link partitions, the cloud
replica simply lags and catches up on reconnection; the control loop is
unaffected.}
\label{fig:repl}
\end{figure}

This section explains how the system spans the edge and the cloud without letting
the cloud become a dependency of control. Each facility's on-premises Kafka
cluster is replicated to an independent cloud cluster by a dedicated mirroring
service (a Kafka MirrorMaker deployment), and a second recorder and further
analytics consume the replica (Fig.~\ref{fig:repl}). In our development
environment we run a customized mirror that forwards only when the destination is
back up, so a cloud or link outage produces a clean catch-up rather than lost or
duplicated data. Looking ahead, we intend the edge and cloud clusters to run
fully independently, coupled only by the mirror, so that neither side can
overload the field layer. The design question is who owns the authoritative state
of a production job. A production job is the period during which a line makes a
particular part or fills a particular order, bounded by a start and an end, much
like a work order or a shift.

Our answer is that the edge is authoritative. The on-premises side creates and
owns job identifiers and line state, such as running or idle, product, and job
start and end, and publishes that state to Kafka. The cloud side is a replica: it
does not invent job identifiers, it reads them from the replicated stream and
aligns its own records to them.

The consequence we care about most is that control does not depend on the cloud.
If the WAN link degrades or drops, the edge line keeps closing its control loop
on local Kafka with no change in behavior; replication lags and catches up when
connectivity returns. Cloud analytics gain a durable, aggregated view across
lines and facilities without ever being on the critical control path. We trade
strong global consistency, since the cloud is eventually consistent with the
edge, for local autonomy and partition tolerance, which is the right trade for a
control system whose correctness must not depend on a remote data center.
Job-completion events additionally notify downstream cloud automation such as
reporting through a managed queue, which keeps batch work off the real-time path.

\section{Latency and Throughput}
\label{sec:latency}

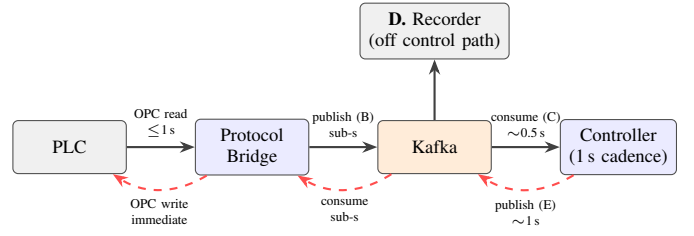
\begin{figure}[t]
\centering
\vspace*{10pt}     
\adjustbox{max width=\columnwidth}{%
\begin{tikzpicture}[font=\scriptsize, node distance=5mm and 9mm]
\node[plantbox] (plc) {PLC};
\node[edgebox, right=of plc] (br) {Protocol\\Bridge};
\node[brokerbox, right=of br] (k) {Kafka};
\node[edgebox, right=of k] (ct) {Controller\\(1\,s cadence)};
\node[edgebox, above=8mm of k, fill=gray!12] (d) {\textbf{D.} Recorder\\(off control path)};
\draw[dataflow] (plc) -- node[above,font=\tiny,align=center]{OPC read\\$\le$1\,s} (br);
\draw[dataflow] (br) -- node[above,font=\tiny,align=center]{publish (B)\\sub-s} (k);
\draw[dataflow] (k) -- node[above,font=\tiny,align=center]{consume (C)\\$\sim$0.5\,s} (ct);
\draw[dataflow] (k) -- (d);
\draw[ctrlflow] (ct) to[bend left=32] node[below,font=\tiny,align=center]{publish (E)\\$\sim$1\,s} (k);
\draw[ctrlflow] (k) to[bend left=32] node[below,font=\tiny,align=center]{consume\\sub-s} (br);
\draw[ctrlflow] (br) to[bend left=32] node[below,font=\tiny,align=center]{OPC write\\immediate} (plc);
\end{tikzpicture}%
}
\caption{Where latency comes from, and the full closed loop. Sense path (solid):
the bridge reads the PLC over OPC-UA, its producer (B) publishes to Kafka, and
the controller's consumer (C) reads. Actuate path (dashed): the controller's
producer (E) publishes the new setpoint to Kafka, the bridge consumes it, and
issues the OPC-UA write back to the PLC. The Recorder (D) reads the same stream
off the control path and is allowed more latency for write throughput. Because
stages pipeline and are paced by the one-second cadence, the round trip is
engineered to 1--2\,s rather than the sum of hops.}
\label{fig:budget}
\end{figure}

This section gives the timing budget of the loop and shows how the system trades
a little latency for throughput without stalling control. The tension is that
controllers want data soon, while the recorder wants to write data efficiently.
The architecture resolves it by batching where it can afford to and reading
promptly where it cannot, so high-rate periods are batched for throughput while
control still sees fresh data. Figure~\ref{fig:budget} places the stages on the
closed loop and shows how each contributes to the budget.

\subsection{The end-to-end timing budget}
\label{sec:budget}
Table~\ref{tab:budget} decomposes the design budget for the full loop, the target
the system is built to. The dominant terms are the one-second sampling and
publication cadence and roughly half-second processing allowances at the bridge
and at each consumer, plus up to a second on the push ingress leg when data comes
from a foreign OPC server. The governing rule is that the loop must never act on
data older than about two seconds. Controllers are scheduled on a one-second
cadence, and on making a decision they write to Kafka, from which the bridge
issues the OPC write-back immediately.

\begin{table}[t]
\caption{Design latency budget for the closed loop. Values are engineered
targets from the deployed configuration, not measured distributions.}
\label{tab:budget}
\centering
\adjustbox{max width=\columnwidth}{%
\begin{tabular}{@{}ll@{}}
\toprule
Stage & Budget \\
\midrule
Foreign OPC to our OPC (push gateway) & $\le 1$\,s \\
Bridge: read change, align time, cache & $\approx 0.5$\,s \\
Sampling / publication cadence & $1$\,s \\
Kafka publish and broker & sub-second \\
Consumer read (time-boxed poll) & $\approx 0.5$\,s \\
Controller decision (scheduled cadence) & $1$\,s \\
Write to Kafka, bridge, PLC & immediate \\
\midrule
\textbf{End-to-end design target} & \textbf{$\sim$1--2\,s} \\
\bottomrule
\end{tabular}%
}
\end{table}

\subsection{Producers, consumers, and the recorder}
On the \textbf{producer} side (B in Fig.~\ref{fig:budget}), the bridge's Kafka
producer coalesces messages into small batches with compression, favoring network
and broker efficiency at the cost of a little latency. On the \textbf{consumer}
side (C), each service reads with bounded internal buffers, so a slow downstream
stage applies back-pressure rather than growing memory without bound, and reads
are time-boxed, so a consumer takes whatever is available within a short timeout
and proceeds rather than blocking; a quiet upstream never stalls a controller.
Per-stage processing time, throughput, queue depth, and liveness are exported as
Prometheus metrics~\cite{prometheus}, which is how we observe and tune the
pipeline. The write-back \textbf{producer} (E) on the controller publishes the
new setpoint back to Kafka the moment a decision is made, so actuation is not
batched. Finally, the \textbf{recorder} (D) reads the same stream off the control
path and writes to storage in batches on a background path, so database writes
never block stream consumption; low-latency dashboards are fed from a metrics
store while the full record lands in a relational time-series
database~\cite{timescale}.

\section{Evaluation}
\label{sec:eval}

We evaluate the fabric on three levels, in decreasing order of directness: a
measured Kafka transport latency taken from the cluster itself; the design budget
of Table~\ref{tab:budget}, derived from the deployed configuration; and
literature-grounded bounds for the remaining components, drawn from published
studies of the same building blocks. We keep these separate so that no
third-party figure is read as a measurement of our own system.

\subsection{Measured Kafka transport latency}
\label{sec:kafkalat}
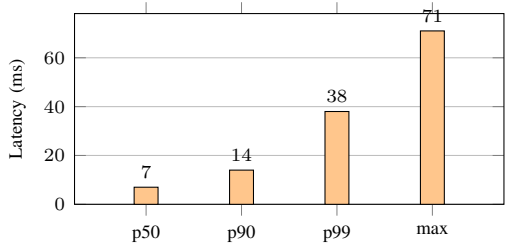
\begin{figure}[t]
\centering
\vspace*{10pt}     
\adjustbox{max width=\columnwidth}{%
\begin{tikzpicture}
\begin{axis}[
    width=0.82\columnwidth, height=4.1cm,
    ybar, bar width=9pt,
    ymin=0,
    ylabel={Latency (ms)},
    symbolic x coords={p50, p90, p99, max},
    xtick=data,
    nodes near coords, nodes near coords style={font=\scriptsize},
    enlarge x limits=0.25,
    ymajorgrids, tick label style={font=\scriptsize},
    label style={font=\scriptsize},
]
\addplot[fill=orange!45] coordinates {(p50,7) (p90,14) (p99,38) (max,71)};
\end{axis}
\end{tikzpicture}%
}
\caption{Measured Kafka transport (produce-to-consume) latency percentiles.
Values are representative of production operation. The scale, milliseconds, is
the point: it is negligible against the one-second sampling cadence, so the
messaging fabric is not the latency-limiting stage.}
\label{fig:kafkalat}
\end{figure}

The one segment we measure directly in production is the \emph{transport latency}
of the fabric: the time from a record being produced to Kafka to its delivery to
a consumer, taken from the Kafka message timestamp and the consumer's receive
time on the same clock. This is the cost of the communication layer proper,
isolated from the sampling cadence and application processing that dominate the
rest of the budget. Figure~\ref{fig:kafkalat} reports the distribution. The
transport latency is on the order of milliseconds, two to three orders of
magnitude below the one-second sampling cadence. The conclusion is decisive: the
broker is not the bottleneck. The end-to-end budget is governed by the physical
sampling cadence and the controller schedule, not by the messaging fabric. This
is the property that justifies using a durable, replayable log as the transport
rather than a lighter-weight bus.

\subsection{Literature-grounded component latency}
\label{sec:litbudget}
For the components we do not benchmark ourselves, independent studies place each
well inside our per-stage budget. Peer-reviewed and industry benchmarks report
Apache Kafka end-to-end latencies in the single- to low-tens-of-milliseconds
range at moderate throughput~\cite{dobbelaere2017,confluentbench,kleppmann2017},
consistent with our own transport-latency measurement. OPC-UA client-server and
PubSub latency has been characterized in the industrial-communications
literature, including real-time operation over TSN~\cite{bruckner2019,profanter2019},
and broad broker and protocol comparisons quantify the throughput, latency, and
fan-out behavior that motivated our choice of Kafka over RabbitMQ and
MQTT~\cite{dobbelaere2017,naik2017,luzuriaga2015}. Because these are third-party
measurements of our building blocks, we use them only to bound the stages we do
not measure, never to claim a figure for our own system.

\subsection{Toward a fuller study}
The platform already exports the per-stage Prometheus series a deeper evaluation
would build on: broker and consume wait, throughput, processing time,
queue-depth (back-pressure) gauges, and resource and liveness gauges. The natural
next steps are mechanical rather than architectural: promoting the
transport-latency measurement to a clock-synchronized end-to-end histogram across
the full sense-to-actuate
path, and characterizing recording throughput, horizontal scaling across lines,
and cloud-replica catch-up after a WAN partition. We deliberately avoid a
head-to-head against the pre-existing plant historian, which we cannot instrument
fairly, so our comparison to alternatives in Section~\ref{sec:discussion} is
qualitative and grounded in the cited literature.

\section{Deployment and Operation}
\label{sec:deploy}

\subsection{The recorder}
\label{sec:recorder}
The recorder is a representative consumer. It tracks the lifecycle of each
production job as it observes line state, labels the recorded data with job
identifiers, records events such as value changes and defects, and in the cloud
role triggers downstream automation on job completion. It
exposes HTTP endpoints for service discovery, health, status, and a soft reset
that re-reads configuration without a process restart, suitable for
container-orchestration probes.

\subsection{Packaging and delivery}
All services are containerized with Docker~\cite{docker} and built through a
cloud CI/CD pipeline that publishes images to a registry and draws private
dependencies from an artifact repository. Separate development, QA, and
production environments are isolated. Services are deployed per line, one recorder
and one of each controller per line, parameterized by a line identifier that
selects the right data streams and configuration. This per-line isolation gives fault
containment, since a problem on one line cannot starve another, and a natural
unit of horizontal scaling across a facility and across facilities.

\subsection{Safety and data-quality gating}
Because the loop acts on streamed data, the system defends against stale or
missing data, which is a communication concern as much as a control one.
Producers emit a periodic heartbeat, on the order of one to two seconds; a
consumer that stops seeing it treats the source as lost. Each reading carries its
source timestamp and quality, and a consumer refuses to act on a value that is
too old, since reacting to two-second-stale data can be worse than not reacting.
If too many values go missing, or the heartbeat lapses, the affected control is
disengaged and the process falls back to the PLC's own safe behavior. These rules
make the near-real-time tier fail safe: the streaming loop is only ever an
optimizing layer on top of hard-real-time interlocks it cannot override.

\subsection{In production}
The platform runs on real production lines, for example extrusion processes,
performing closed-loop auto-correction of machine settings, on-line defect
detection, and ML-driven virtual sensing. Operator-facing dashboards are fed from
the metrics store, and edge access is mediated by an identity-aware gateway. The
shared communication libraries are versioned and consumed across many internal
services, which is itself evidence that the single-stream, standard-format
discipline pays off: teams build new consumers against a stable contract.

\subsection{Lessons learned}
Three lessons stand out from operating the fabric. First, \emph{the transport is
rarely the latency limit}. As Section~\ref{sec:eval} shows, the broker adds
milliseconds while the physical sampling cadence and controller schedule add
seconds, so engineering effort is better spent on sampling and scheduling than on
squeezing the broker. Second, \emph{putting time-stamping at the source is worth
the effort}. Once every sample is stamped in the OPC layer, the rest of the
pipeline can batch, replay, and reorder data freely without corrupting its
meaning, which is what makes the high-rate array transport and consumer replay
safe. Third, \emph{the durable log pays for itself in operations, not just in
throughput}. Being able to rewind the stream turned out to matter most for
debugging: a misbehaving controller can be replayed offline against the exact
data it saw, and every commanded setpoint is auditable after the fact, which is
difficult with a broker that forgets a message once delivered.

\subsection{Security and access}
Although security is not the focus of this paper, the fabric inherits several
properties relevant to a zero-trust posture. OPC-UA sessions use encrypted
transport with certificate- or password-based authentication, edge access is
mediated by an identity-aware gateway rather than exposed directly, and per-line
isolation limits the blast radius of any single compromised service. Because the
bridge is the only component that writes to the PLC, actuation has one auditable
choke point to monitor and harden.

\section{Discussion}
\label{sec:discussion}

The central claim of this paper is that the value is architectural, so this
section states plainly what is standard, what is novel, and why the standard
pieces were chosen.

\subsection{Standard components, novel composition}
Every protocol and broker we use is mainstream: Kafka, OPC-UA, JSON, a relational
time-series database, and a metrics store. What is novel, and under-reported in
the literature for \emph{closed-loop} industrial control, is the composition: a
single per-line stream that serves control, monitoring, ML, and recording at
once; a bridge that makes that stream the sole conduit for both sensing and
actuation; a transport that carries high-rate signals over a coarse cadence
without losing time fidelity; and an edge-authoritative replication model that
buys WAN-resilient autonomy without giving up cloud analytics. Each piece is
simple; the value is in stitching them together and showing that the result meets
near-real-time control requirements in production.

\subsection{Why Kafka rather than OPC-UA fan-out, RabbitMQ, or MQTT}
The broker choice was deliberate. OPC-UA is excellent as a device protocol but is
not a shared analytics bus: a single OPC server slows down as many clients
subscribe to it, so using it as the fan-out layer would couple every consumer to
the health of the PLC-facing server. Kafka instead becomes the single source of
truth that any number of services read independently. Its pull-based subscription
model lets services read at their own pace and stay loosely coupled, so one slow
or restarting consumer cannot disturb the others, and its durable, replayable log
enables consumer warm-up, controller back-testing, and forensic auditing of every
commanded setpoint. We considered RabbitMQ, a capable low-latency broker, but its
push-to-subscriber model becomes a bottleneck as clients grow and gives no
replayable log; pushing as soon as data arrives removes the consumer's ability to
pace itself, which is the control we want at the fan-out layer. MQTT is
attractively lightweight but reintroduces the very problems we solve at the
bridge, such as timestamp unification, and again lacks a durable log. Our design
uses push where push is
right, at the edge ingress gateway that delivers fresh data the instant it
changes, and pull where pull is right, at the Kafka fan-out where consumers must
not be overwhelmed. Table~\ref{tab:brokers} summarizes the comparison against the
properties that decided the choice.

\begin{table}[t]
\caption{Messaging substrates against the properties that mattered for a
single-stream, many-consumer control fabric.}
\label{tab:brokers}
\centering
\footnotesize
\adjustbox{max width=\columnwidth}{%
\begin{tabular}{@{}lcccc@{}}
\toprule
Property & OPC-UA & MQTT & RabbitMQ & Kafka \\
\midrule
Durable, replayable log      & no      & no      & no      & \textbf{yes} \\
Consumer sets its own pace   & no      & no      & no      & \textbf{yes} \\
Fan-out to many consumers    & limited & yes     & limited & \textbf{yes} \\
Ordered delivery             & n/a     & no      & partial & \textbf{yes} \\
Source-timestamped by design & yes     & no      & no      & via bridge \\
\bottomrule
\end{tabular}%
}
\end{table}

\subsection{Soft, not hard, real time}
This is the most important caveat for a real-time-communication audience. The
Kafka loop operates at sub-second-to-seconds granularity and makes no hard timing
guarantee; the deterministic, safety-critical control stays in the PLC. Our claim
is that a large and valuable class of supervisory, optimizing control can live at
this softer tier, and that standard streaming infrastructure fits it well.

\section{Limitations and Future Work}

The system is near-real-time by construction; we do not bound worst-case latency,
and our direct measurement covers the broker-transport stage rather than the full
sense-to-actuate loop. JSON serialization trades bytes for legibility and would
not scale to very high-frequency signals. The per-line broker simplifies
isolation but multiplies operational surface area across a large facility. Future
work includes a clock-synchronized end-to-end latency histogram, a
latency-throughput study across batching parameters, evaluation of a binary
schema format for high-rate signals, exactly-once semantics on the write-back
path, and a formal statement of the edge-cloud consistency model.

\section{Conclusion}

We described a production communication fabric that brings near-real-time,
closed-loop control to the manufacturing floor using only standard streaming and
industrial components. By publishing each line's equipment data once to a durable
Kafka log and letting control, monitoring, ML, and recording consume it
independently, by bridging to PLCs through a single OPC-UA gateway that also
aligns timestamps and carries high-rate signals, and by replicating to the cloud
under an edge-authoritative model, the system achieves decoupling, observability,
and WAN-resilient autonomy without a custom protocol. The contribution is
architectural and experiential: a reusable pattern for composing real-time
communication infrastructure with industrial IoT and ML, and an honest account of
where it provides soft rather than hard real-time guarantees.

\section*{Acknowledgment}
The authors thank the engineering and data-science teams that built and operate
the platform described here.



\begin{thebibliography}{00}

\bibitem{lasi2014} H.~Lasi, P.~Fettke, H.-G.~Kemper, T.~Feld, and M.~Hoffmann,
``Industry 4.0,'' \emph{Business \& Information Systems Engineering}, vol.~6,
no.~4, pp.~239--242, 2014.

\bibitem{lee2015} J.~Lee, B.~Bagheri, and H.-A.~Kao, ``A cyber-physical systems
architecture for Industry 4.0-based manufacturing systems,'' \emph{Manufacturing
Letters}, vol.~3, pp.~18--23, 2015.

\bibitem{zhong2017} R.~Y.~Zhong, X.~Xu, E.~Klotz, and S.~T.~Newman, ``Intelligent
manufacturing in the context of Industry 4.0: A review,'' \emph{Engineering},
vol.~3, no.~5, pp.~616--630, 2017.

\bibitem{icad2026} Z.~Gu, J.~E.~Hernandez, T.~Cook, J.~Burtenshaw, S.~Scott, and
C.~Couch, ``Benchmarking hybrid deep learning architectures for predictive
maintenance in Industry 4.0,'' in \emph{Proc. IEEE Int. Conf. on AI and Data
Analytics (ICAD)}, 2026.

\bibitem{tao2019} F.~Tao, H.~Zhang, A.~Liu, and A.~Y.~C.~Nee, ``Digital twin in
industry: State-of-the-art,'' \emph{IEEE Trans. Ind. Informat.}, vol.~15, no.~4,
pp.~2405--2415, 2019.

\bibitem{wollschlaeger2017} M.~Wollschlaeger, T.~Sauter, and J.~Jasperneite,
``The future of industrial communication: Automation networks in the era of the
Internet of Things and Industry 4.0,'' \emph{IEEE Ind. Electron. Mag.}, vol.~11,
no.~1, pp.~17--27, 2017.

\bibitem{vitturi2019} S.~Vitturi, C.~Zunino, and T.~Sauter, ``Industrial
communication systems and their future challenges: Next-generation Ethernet,
IIoT, and 5G,'' \emph{Proc. IEEE}, vol.~107, no.~6, pp.~944--961, 2019.

\bibitem{shi2016} W.~Shi, J.~Cao, Q.~Zhang, Y.~Li, and L.~Xu, ``Edge computing:
Vision and challenges,'' \emph{IEEE Internet Things J.}, vol.~3, no.~5,
pp.~637--646, 2016.

\bibitem{satya2017} M.~Satyanarayanan, ``The emergence of edge computing,''
\emph{Computer}, vol.~50, no.~1, pp.~30--39, 2017.

\bibitem{opcua} International Electrotechnical Commission, ``IEC~62541: OPC
Unified Architecture,'' IEC, Geneva, Switzerland.

\bibitem{profanter2019} S.~Profanter, A.~Tekat, K.~Dorofeev, M.~Rickert, and
A.~Knoll, ``OPC UA versus ROS, DDS, and MQTT: Performance evaluation of
industrial protocols,'' in \emph{Proc. IEEE Int. Conf. Industrial Technology
(ICIT)}, 2019, pp.~955--962.

\bibitem{tsn} IEEE, ``IEEE~802.1Q: Bridges and bridged networks (Time-Sensitive
Networking),'' IEEE Standards Association.

\bibitem{bruckner2019} D.~Bruckner \emph{et al.}, ``An introduction to OPC UA TSN
for industrial communication systems,'' \emph{Proc. IEEE}, vol.~107, no.~6,
pp.~1121--1131, 2019.

\bibitem{kafka} J.~Kreps, N.~Narkhede, and J.~Rao, ``Kafka: A distributed
messaging system for log processing,'' in \emph{Proc. NetDB}, 2011.

\bibitem{kleppmann2017} M.~Kleppmann, \emph{Designing Data-Intensive
Applications}. Sebastopol, CA, USA: O'Reilly Media, 2017.

\bibitem{akidau2015} T.~Akidau \emph{et al.}, ``The dataflow model: A practical
approach to balancing correctness, latency, and cost in massive-scale, unbounded,
out-of-order data processing,'' \emph{Proc. VLDB Endowment}, vol.~8, no.~12,
pp.~1792--1803, 2015.

\bibitem{carbone2015} P.~Carbone \emph{et al.}, ``Apache Flink: Stream and batch
processing in a single engine,'' \emph{IEEE Data Eng. Bull.}, vol.~38, no.~4,
pp.~28--38, 2015.

\bibitem{amqp} OASIS, ``Advanced Message Queuing Protocol (AMQP) Version 1.0,''
OASIS Standard, 2012; see also ISO/IEC~19464.

\bibitem{mqtt} OASIS, ``MQTT Version 5.0,'' OASIS Standard, 2019; see also
ISO/IEC~20922.

\bibitem{dobbelaere2017} P.~Dobbelaere and K.~S.~Esmaili, ``Kafka versus
RabbitMQ: A comparative study of two industry reference publish/subscribe
implementations,'' in \emph{Proc. 11th ACM Int. Conf. Distributed and
Event-Based Systems (DEBS)}, 2017, pp.~227--238.

\bibitem{naik2017} N.~Naik, ``Choice of effective messaging protocols for IoT
systems: MQTT, CoAP, AMQP and HTTP,'' in \emph{Proc. IEEE Int. Systems Eng. Symp.
(ISSE)}, 2017, pp.~1--7.

\bibitem{luzuriaga2015} J.~E.~Luzuriaga \emph{et al.}, ``A comparative evaluation
of AMQP and MQTT protocols over unstable and mobile networks,'' in \emph{Proc.
IEEE Consumer Communications and Networking Conf. (CCNC)}, 2015, pp.~931--936.

\bibitem{isa95} International Society of Automation, ``ANSI/ISA-95: Enterprise-Control System Integration,'' ISA, USA.

\bibitem{cap} S.~Gilbert and N.~Lynch, ``Brewer's conjecture and the feasibility
of consistent, available, partition-tolerant web services,'' \emph{ACM SIGACT
News}, vol.~33, no.~2, pp.~51--59, 2002.

\bibitem{ntp} D.~Mills, J.~Martin, J.~Burbank, and W.~Kasch, ``Network Time
Protocol Version 4: Protocol and algorithms specification,'' RFC~5905, IETF,
2010.

\bibitem{ptp} IEEE, ``IEEE~1588: Standard for a precision clock synchronization
protocol for networked measurement and control systems,'' IEEE Standards
Association, 2019.

\bibitem{prometheus} B.~Brazil, \emph{Prometheus: Up \& Running}. Sebastopol,
CA, USA: O'Reilly Media, 2018.

\bibitem{timescale} Timescale, Inc., ``TimescaleDB: An open-source time-series
database built on PostgreSQL,'' technical documentation.

\bibitem{docker} D.~Merkel, ``Docker: Lightweight Linux containers for
consistent development and deployment,'' \emph{Linux Journal}, vol.~2014,
no.~239, 2014.

\bibitem{confluentbench} Confluent, Inc., ``Benchmarking Apache Kafka, Apache
Pulsar, and RabbitMQ: Which is the fastest?,'' technical report, 2020.
[Online; industry benchmark, not peer-reviewed.]

\end{thebibliography}
\end{document}